\documentclass{article}

\usepackage{microtype}
\usepackage{graphicx}
\usepackage{subfigure}
\usepackage{booktabs} 

\usepackage{hyperref}

\usepackage[accepted]{icml2018}

\icmltitlerunning{Singing Style Transfer Using Cycle-Consistent Boundary Equilibrium Generative Adversarial Networks}

\begin{document}

\twocolumn[
\icmltitle{Singing Style Transfer Using Cycle-Consistent Boundary Equilibrium Generative Adversarial Networks}



\icmlsetsymbol{equal}{*}

\begin{icmlauthorlist}
\icmlauthor{Cheng-Wei Wu}{ntu,as}
\icmlauthor{Jen-Yu Liu}{ntu,as}
\icmlauthor{Yi-Hsuan Yang}{as}
\icmlauthor{Jyh-Shing R. Jang}{ntu}
\end{icmlauthorlist}

\icmlaffiliation{ntu}{National Taiwan University;}
\icmlaffiliation{as}{Academia Sinica, Taiwan;}

\icmlcorrespondingauthor{Cheng-Wei Wu}{haley.wu@mirlab.org}
\icmlkeywords{Machine Learning, Generative Adversarial Networks, Singing Style Transfer, ICML}

\vskip 0.3in
]



\printAffiliationsAndNotice{}  

\begin{abstract}
	Can we make a famous rap singer like Eminem 
sing whatever our favorite song?
	Singing style transfer attempts to make this possible, by replacing the vocal of a song from the source singer to the target singer.
    This paper presents a method that learns from unpaired data for singing style transfer using generative adversarial networks.
\end{abstract}

\section{Introduction}
\begin{figure}
		\begin{center}
      		\centerline{\includegraphics[width=\columnwidth]{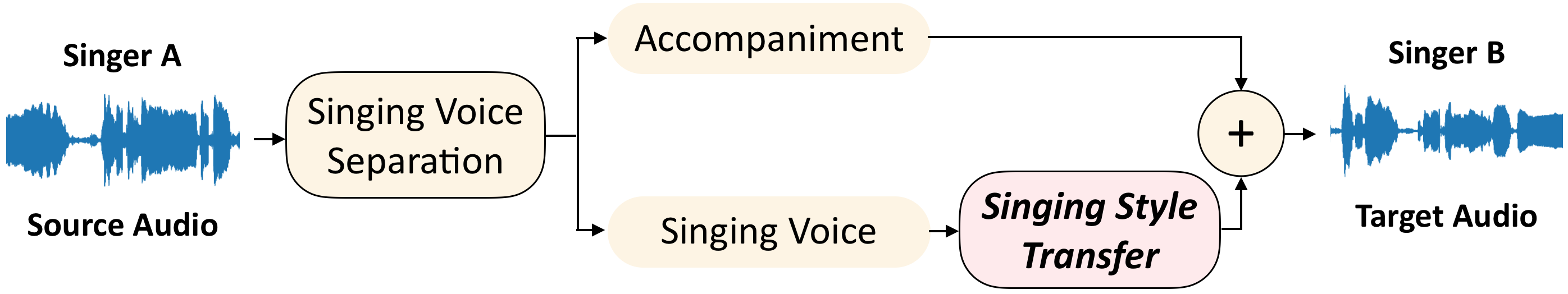}}
      		\caption{System framework for singing style transfer.}
      		\label{fig:framework}
		\end{center}
	\vskip -0.4in
	\end{figure}

	Figure \ref{fig:framework} shows a two-stage framework combining singing voice separation with singing style transfer.
    The audio from a \emph{source} singer is separated into accompaniment and vocal first, and then the singing style of the separated vocal is changed to that of a \emph{target} singer.
    Finally, the separated accompaniment is integrated with the style-transferred vocal.
    Our focus here is on the singing style transfer model. 
    We assume that the input to this model is clean, meaning that it has been perfectly separated from the source audio.

    We evaluate the proposed model through a listening test under the \emph{inside test} scenario, meaning that the training and test sets have overlaps. We show audio result for both inside and outside tests on our project website: \url{http://mirlab.org/users/haley.wu/cybegan/}.


\section{Methodology}
	The method proposed by Gatys \emph{et al.} \yrcite{gatys16cvpr} leads to the earliest successful examples of image style transfer.
    The method has been extended to perform audio style transfer, either to transfer sound texture \cite{ulyanov2016} or to stylize an acapella cover to match the original vocal \cite{bohan2017}.
	However, the method requires the availability of paired data for training (i.e., the source and target singers need to have sung the same songs), which is not readily available for arbitrary target singers.

    In this paper, we propose to use a CycleGAN-CNN \cite{CycleGAN2017} based model due to its success in image style transfer without using paired data.
	We use log-magnitude spectrogram as the model input, with 1,024-sample, 1/4 overlapping windows for the STFT for songs sampled at 44.1kHz.
	To capture information in the spectra, we view a $T\times F$ 2D spectrogram as a $T \times 1$ image with $F$ channels, and use 1D convolution \cite{jy18ss} for all the layers in our model.
	Since a song may have variable length, we adopt a fully-convolutional design \cite{long2015fully,liu16mm} and use no pooling layers.
	To get the time-domain audio signal from the modified spectrogram, we use the classic Griffin-Lim algorithm \yrcite{griffin1984signal}.

	\begin{figure*}
	\vskip 0.2in
		\begin{center}
			\centerline{\includegraphics[width=\textwidth]{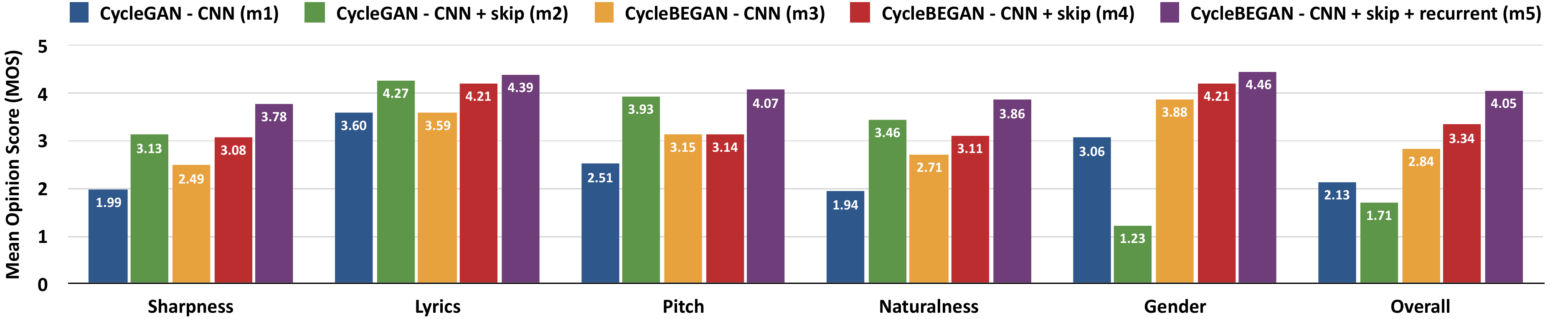}}
			\caption{Results of subjective listening test: mean opinion scores (MOS) in six performance metrics.}
			\label{fig:subjective_mos}
		\end{center}
	\vskip -0.2in
	\end{figure*}

	We experiment with a few extensions of this basic model.
    First, we use a U-Net architecture \cite{ronneberger2015u} and add symmetric \emph{skip connections} between the encoding and decoding layers to preserve the details of the given source audio and to increase the sharpness of the vocal.
	Second, we use BEGAN \cite{berthelot2017began} instead of GAN in an attempt to stabilize the training process.
    Moreover, in this \emph{CycleBEGAN} design, we replace the original PatchGAN-based discriminator \cite{li2016precomputed,ledig2017photo,pix2pix2017} with an auto-encoder having identical architecture with the generator.
	Finally, being inspired by \cite{jy18ss}, we add a \emph{recurrent layer} (using GRU \cite{cho-al-emnlp14}) before the output convolution layer to account for the temporal dependency in music.

\section{Experiments}
	We train our model on the iKala dataset \cite{chan2015vocal}, which comes with 252 30-second excerpts of clean vocal recordings. We cut the excerpts into 5-second clips, with 4-second overlaps.
    We randomly select 2,800 segments from each gender as the training set, and 100 segments from each gender as the test set.
    As part of each test clip may have been observed in the training set, this is an inside test.

    To simplify the problem, we train our model to perform ``gender transfer'', i.e., either male-to-female or female-to-male.
    We implement and compare in total five different models (denoted as m1--m5 in Figure~\ref{fig:subjective_mos}), including the basic `CycleGAN-CNN' and the most sophisticated `CycleBEGAN-CNN+skip+recurrent.'

	We conduct a subjective listening test to evaluate the performance of style transfer.
    The subjects are asked to listen to the source and transferred audio for six 15-second clips, including 3 clips for male-to-female and 3 clips for female-to-male.
	A subject has to compare the result of two randomly chosen models for each set, and evaluate the results in terms of the following six metrics, using a five-point Likert scale:
	\begin{itemize}
	\itemsep0em
		\item The \texttt{sharpness} of the transferred singing voice.
		\item Whether the \texttt{lyrics} (main content of the song) is intelligible and is consistent with that in the source.
		\item The perceived \texttt{pitch} accuracy (relative pitch) of the transferred singing voice.
		\item \texttt{Naturalness}: whether the  transferred singing voice is generated by human not by machine.
		\item Whether the \texttt{gender} of the singer has been changed.
		\item \texttt{Overall}: whether the transferred  voice achieve the goal of style transfer and its overall perceptual quality.
	\end{itemize}

\begin{figure}
		\begin{center}
			\centerline{\includegraphics[width=\columnwidth]{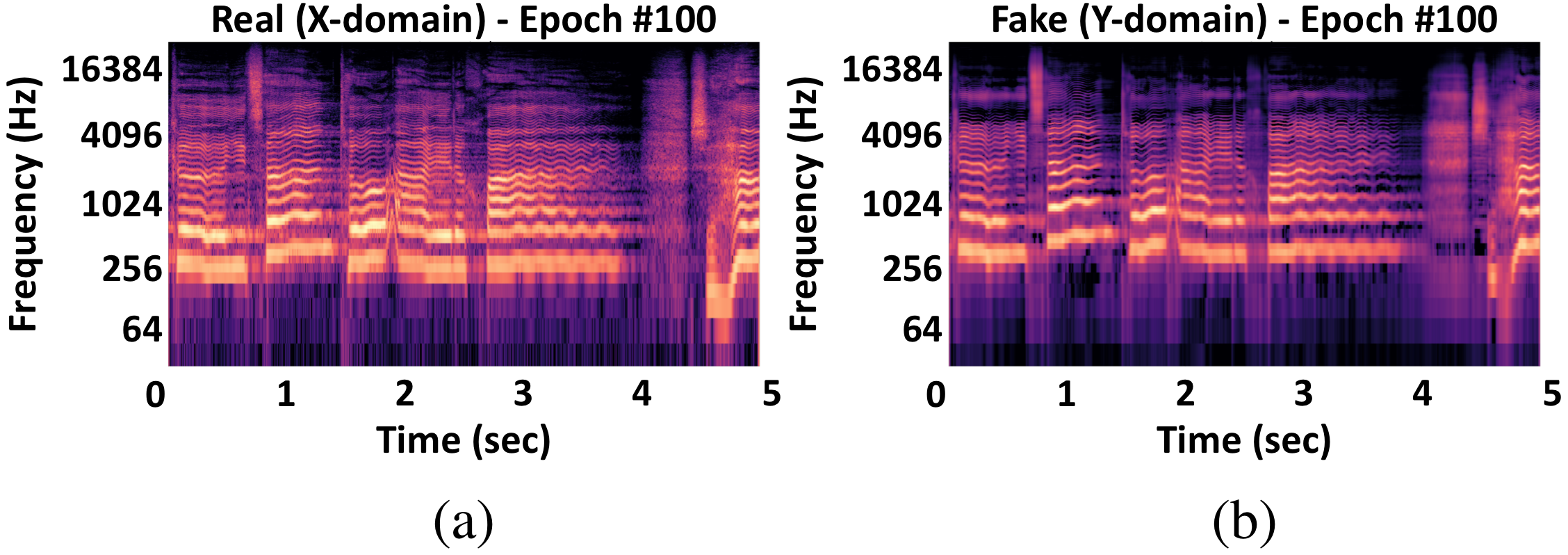}}
			\caption{The spectrograms of the (a) source audio sung by a male singer and the (b) style transferred audio (male-to-female) of an example song, using CycleBEGAN-CNN+skip+recurrent (m5). We can see that the pitches in (b) are higher than those in (a).}
			\label{fig:exp_result}
		\end{center}
	\vskip -0.2in
	\end{figure}

	Figure~\ref{fig:subjective_mos} shows the mean opinion scores (MOS) in these six metrics from 69 adults (19 females).
    The following observations are made.
	First, in terms of \texttt{sharpness} and \texttt{lyrics}, all the models with symmetric skip connections (m2, m4, m5) outperform those without them (m1, m3).
    This shows that the audio details propagated via the skip connections from the encoder to the decoder help generate high-resolution audio.
	Second, in terms of \texttt{pitch}, the CycleBEGANs (m3, m4, m5) outperform the simple CycleGAN model (m1), but not the CycleGAN with skip connections (m2).
    By listening to the result, we found that m2 works like a normal auto-encoder which only reconstructs the input without transferring the  style.
    Therefore, it preserves the pitch well but performs very poorly in \texttt{gender}.
	Next, in terms of \texttt{naturalness}, the outputs of the models with skip connections (m2, m4, m5) score higher than the others (m1, m3) as expected.
    Similar to \texttt{pitch}, the outputs of m2 sound more natural than the outputs from the other models,  except for m5.
	Finally, the scores in \texttt{gender} is consistent with the result of the \texttt{overall} performance.
    This is expected, because the goal is to transfer the singing style male$\rightarrow$female or female$\rightarrow$male.
    CycleBEGAN models (m3, m4, m5) again outperform the CycleGAN ones (m1, m2), which verifies that the the integration of CycleGAN with BEGAN effectively improves the result of style transfer.
	The most sophisticated model m5 outperforms the other models in all the metrics, reaching scores that are \textbf{close to or over 4} on the five-point Likert scale.
	 Figure \ref{fig:exp_result} shows an example of the transfered spectrograms by m5.

\section{Conclusion and Future Work}
	We have proposed a new approach for singing style transfer without paired data by combining CycleGAN with the training strategy of BEGAN.
	We have shown that symmetric skip connections increase the sharpness of the result, and that BEGAN plays an important role in transferring the singer properties.
	Future work will be done to transfer singer identities instead of only gender, and to include a singing voice separation model \cite{jansson17ismir,jy18ss}
    to deal with real songs instead of clean vocals.
\nocite{langley00}

\bibliography{references}
\bibliographystyle{icml2018}





\end{document}